\documentclass[letterpaper, 10 pt, conference]{IEEEtran}

\usepackage{lipsum}
\usepackage{graphicx}
\usepackage{amsmath}
\usepackage{xcolor}

\usepackage{hyperref}
\usepackage[capitalize]{cleveref}
\usepackage{amssymb}
\usepackage{cite}

\graphicspath{ {./Resources/} }

\title{\LARGE \bf Data-driven classification of low-power communication signals by an unauthenticated user using a software-defined radio}

\author{
Tarun Rao Keshabhoina$^{1}$ and Marcos M. Vasconcelos$^{2}$

\thanks{$^{1}$T. R. Keshabhoina is with the Department of Electrical Engineering and the Commonwealth Cyber Initiative, Virginia Tech, USA. Email:
    {\tt tarunrao@vt.edu}}
\thanks{$^{2}$M. M. Vasconcelos is with the Department of Electrical and Computer Engineering, FAMU-FSU College of Engineering,
    Florida State University, USA. Email:
    {\tt  m.vasconcelos@fsu.edu}}}
        
\IEEEoverridecommandlockouts    

\begin{document}

\maketitle

\begin{abstract}
    Many large-scale distributed multi-agent systems exchange information over low-power communication networks. In particular, agents intermittently communicate state and control signals in robotic network applications, often with limited power over an unlicensed spectrum, prone to eavesdropping and denial-of-service attacks. In this paper, we argue that a widely popular low-power communication protocol known as \textit{LoRa} is vulnerable to denial-of-service attacks by an unauthenticated attacker if it can successfully identify a target signal's bandwidth and spreading factor. Leveraging a structural pattern in the LoRa signal's instantaneous frequency representation, we relate the problem of jointly inferring the two unknown parameters to a  classification problem, which can be  efficiently implemented using neural networks.
\end{abstract}
    
\section{Introduction}

    Multi-agent robotic systems are used in various modern applications, including industrial automation, agriculture, and environmental monitoring~\cite{ismail2018survey, rasheed2022review}. In these systems, autonomous robots work together to accomplish a common goal, such as monitoring an environment or cooperatively completing a task. In such systems, coordination, and communication among the robots are critical to their success. Each robot must be aware of the state and actions of the other robots in the system to coordinate their actions and achieve their goals. For example, in an agricultural monitoring system, each robot may be responsible for monitoring a different field area, and they must coordinate their movements to ensure that the entire field is covered. Therefore, communication among the robots must be reliable, even in challenging scenarios such as remote or outdoor environments, which are subject to disruption by obstacles or malicious interference. Protecting such networks against denial-of-service attacks is of paramount importance to prevent service disruption and economic loss.
    
    \begin{figure}[ht]
        \centering
        \includegraphics[width=0.8\columnwidth]{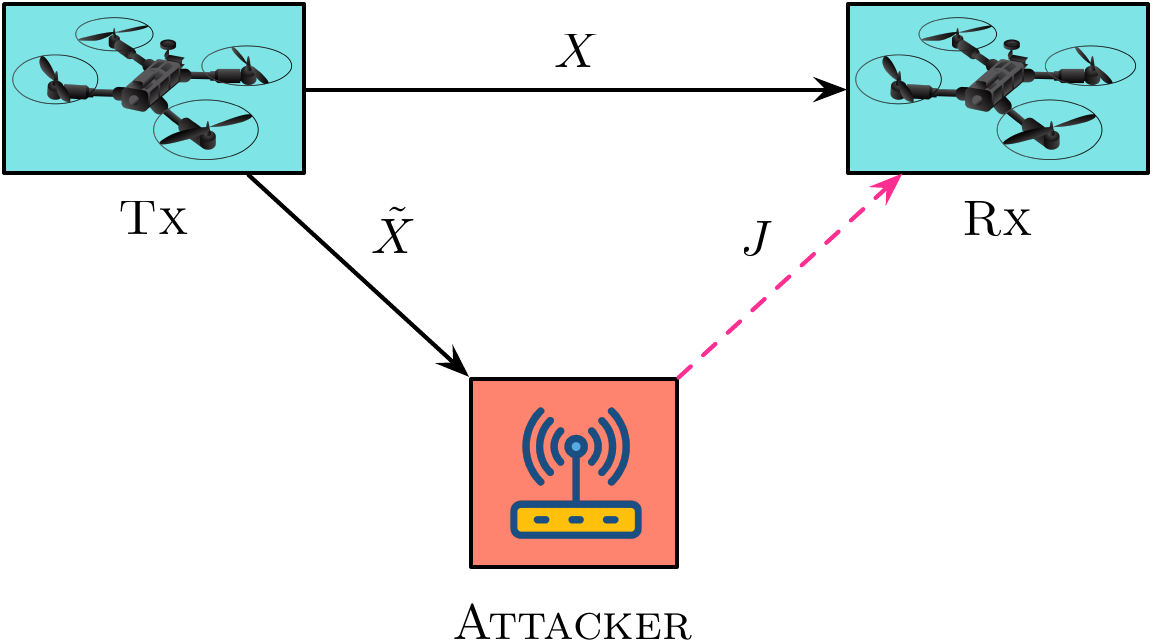}
        \caption{Block diagram for the communication scenario herein: two legitimate agents communicate a signal represented by $X$, an attacker observes a correlated signal $\tilde{X}$, with the intent to emit a jamming signal $J$.}
        \label{fig:jamming}
    \end{figure}
    
    A Low Power Wide Area Network (LPWAN) protocol LoRaWAN (Long Range Wide Area Network) offers long-range and low-power communication capabilities well-suited to multi-agent robotic systems~\cite{adelantado2017understanding}. Additionally, LoRaWAN supports creating large-scale networks with multiple nodes, making it an ideal solution for coordinating the activities of large groups of robots communicating intermittently. While LoRaWAN is one of the most robust and resilient low-power communication protocols, it is still vulnerable to a class of denial-of-service attacks known as \textit{jamming}. 
      
    A jamming attack follows the diagram in \cref{fig:jamming}: a transmitting agent, Tx, sends a signal to a receiving agent, Rx; the transmitted signal is intercepted by an attacker using a software-defined radio unit; The attacker then infers two private parameters used for communication between Tx and Rx, and subsequently sends a jamming signal to interfere with the transmitted signal at the receiver.

    \subsection{Related Work}
        Wireless communication protocols transmit over the air, which makes them vulnerable to interference from any radio transmitter within their vicinity. This fundamental aspect of shared media in wireless networks has made way for extensive research in the wireless jamming domain~\cite{zou2016survey, xu2005feasibility, liu2016physical}. Energy-constrained jamming methodologies attempt to block the channel in reaction to transmission activity to save power. Herein, we discuss such a reactive jamming strategy for LoRa PHY. Securing communication systems and improving  performance in the presence of \textit{intelligent} jammers~\cite{zhang2022robust, zhang2023robust} is the motivation to this work. 

        Numerous studies have examined the throughput and performance of ultra-narrow band (UNB) and spread spectrum-based technologies in the unlicensed Industrial, Scientific, and Medical (ISM) band~\cite{reynders2016range, naik2018lpwan}. Amongst these, a comprehensive study of PHY layer vulnerabilities, countermeasures and security features of LoRaWAN are presented in~\cite{ruotsalainen2022lorawan}, and its authors also provide a brief overview of jamming methodologies for LoRa. Long-range transmissions on LoRa are susceptible to several attack strategies such as replay attacks, wormhole attacks, and compromising network key information, in addition to jamming~\cite{ruotsalainen2022lorawan}.
        
        LoRa's medium access control (MAC) layer design introduces many configurable parameters that affect its service reliability. An in-depth explanation of such parameters, and their resulting performance tradeoffs are presented in~\cite{magrin2021configurable}. Choices of these parameters, driven by service requirements, also play a role in the PHY layer encoding of signals, having implications on the approaches adopted by intelligent jammers. 
            
        When signals from one packet are $6\mathrm{dB}$ stronger than another, it goes on to be demodulated, leaving the weaker packet to be discarded (this is the so-called \textit{channel capture effect})~\cite{rahmadhani2018lorawan}. Building on the concept, the authors of~\cite{aras2017selective} have shown that LoRa can be jammed using commercially available hardware. Herein, they induce collisions on the channel, by flooding it with numerous packets of identical parameter choices. A more advanced technique, targetting the symbol demodulation process in LoRa was explored in~\cite{hou2021jamming}, introducing the idea of jamming chirps. They revealed that LoRa receivers cannot distinguish between a well-synchronized jamming chirp and a legitmate chirp. 
        
        LoRa was found vulnerable to interference when two packets employ the same configuration of two parameters known as the Bandwidth ($BW$) and Spreading Factor ($SF$). The symbol demodulation process in LoRa involves two steps: first, dechirping, and then, FFT (Fast Fourier Transform). Symbols are determined by identifying peaks within the FFT. When interfering packets utilize the same $BW$ and $SF$, this can cause multiple indiscernible peaks in the FFT, leading to symbol errors~\cite{goursaud2015dedicated}. 
            
        Contemporary work in LoRa jamming exploit this property, and an empirical analysis of the approach is discussed in~\cite{hou2023jamming}. While they prove the effectiveness of this strategy, they also make a hard assumption. Particularly, that the jammer has apriori knowledge of the target signal's $BW$ and $SF$ choices, neccessary for generating the jamming chirps. However, these parameters are generally not available to adversarial agents, which are unathenticated users of the network. 
        
        In this paper, we take a step further, exploring how an adversary may employ a simple neural network implementation to estimate this information and jam LoRa signals reactively, without such assumed knowledge. We provide numerical results on the detection and identification of the $BW$ and $SF$ parameters from observed signals. Then, we quantify the robustness of our model by evaluation on a wide range of signal-to-noise ratio (SNR) levels of signals.

        The rest of this paper is organized as follows. Section II introduces LoRa PHY and the chirp spread spectrum. Section III describes system architecture. Section IV describes our proposed feature extraction technique. Section V describes the architecture of the neural network classifier. Section VI presents our simulation results and discusses our system's performance. Finally, Section VII concludes the paper and outlines future research directions.

    
    \section{Signal description}

        \begin{figure*}[!htb]
        \centering
        \includegraphics[width=\textwidth]{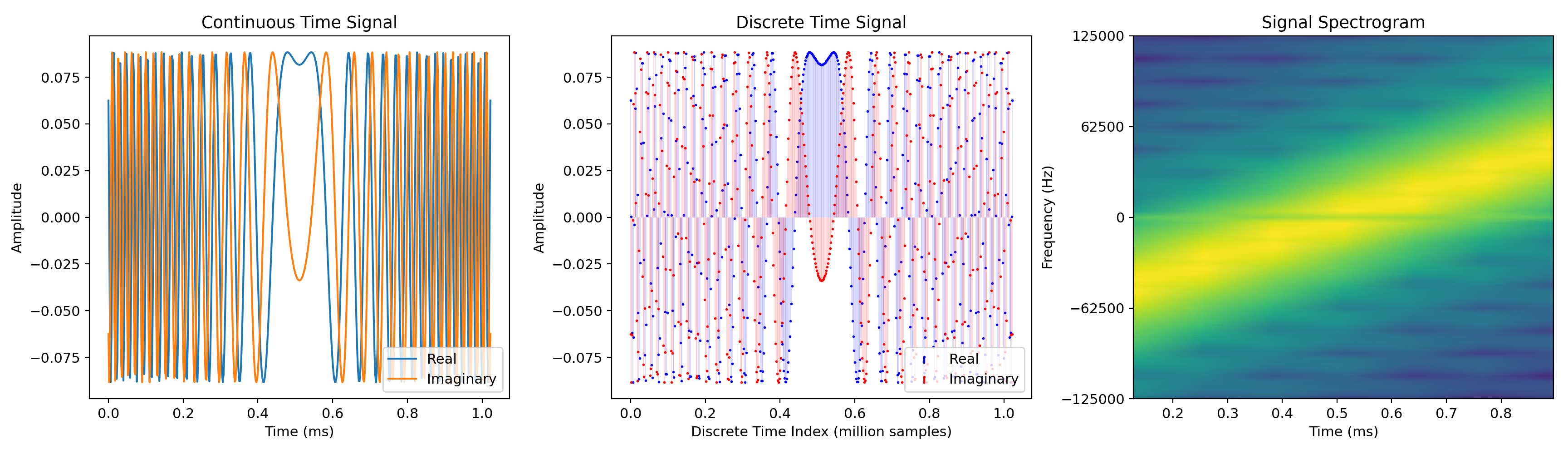}
        \caption{A chirp signal with BW = 125 KHz, and SF = 7 in continuous time (left),  discrete time (middle), and  its spectrogram (right).}
        \label{fig:chirp}
        \end{figure*}
        LoRa PHY is a pass band modulation technique that uses chirp spread spectrum (CSS) to modulate digital information onto a carrier wave. In CSS, a chirp is a signal whose instantaneous frequency increases or decreases linearly as a function of time. 
        
        In LoRa, each transmitted symbol is mapped into a chirp. The bandwidth ($BW$) and spreading factor ($SF$) are the most critical parameters defining a LoRa chirp. The $BW$ corresponds to the range of frequencies of the channel occupied by the chirp, and the $SF$  determines the number of bits transmitted in a symbol. Each symbol carries $SF$ bits (i.e., values ranging from $0$ to $2^{SF} - 1$). The joint choice of $SF$ and $BW$ determines the data rate of the communication link. Following \cite{vangelista2017frequency}, in this section, we describe the CSS modulation. 


    

        A fundamental characteristic of the LoRa chirp is its cyclically shifted frequency. Wherein, the frequency incrementally rises from the initial frequency in discrete steps. Upon reaching the highest frequency, it wraps around to the lowest frequency and continues its ascent until it cycles back to the initial frequency. The chirp encodes information by adjusting its starting frequency according to its symbol value, $s_n$.
        
        Consider the transmission of a sequence of symbols
        $\mathbf{s} := \{s_n\}$. Each symbol carries $SF$ bits, denoted by a vector $\mathbf{w}_n=(w_{n,0},\ldots, w_{n,SF-1})$, where $w_{n,b}\in\{0,1\}$, $b\in\{0,\ldots, SF-1\}$. A new symbol is transmitted every $T_s$ seconds, corresponding to a chirp signal's duration in time.         The value of the symbol $s_n$ is given by
        \begin{equation}
            s_n = \sum_{b=0}^{SF-1}w_{n,b}\times 2^b.
        \end{equation}

 Since $s_n$ can take on $2^{SF}$ distinct values, the channel bandwidth is divided into $2^{SF}$ discrete levels. Each of these levels signifies the starting frequency for a specific symbol value. 
        
        Therefore, the chirp completes $2^{SF}$ discrete steps throughout its duration, in cycling back to its initial frequency. For a chosen bandwidth, $BW$, each step lasts for a duration of $T = 1/BW$ seconds, adding up to the entire symbol duration $T_s$. Thus, $SF$ determines the number of steps, and $BW$ determines the time period of each step, collectively defining the symbol duration, $T_s=2^{SF}/BW$.
        
     
        Let $f_c$ denote the channel's center frequency. The $n$-th transmitted symbol, $s_n$, is mapped into a chirp signal $c_n(t)\in \mathbb{C}$ given by
       \begin{equation}
            c_n(t) = \frac{1}{\sqrt{2^{SF}}}\exp\big\{j\big(2\pi f_n(t)\big)t\big\}, \ \ t \in [0,T_s]
        \end{equation}
        where, 
        \begin{equation}
            f_n(t) = f_c+\mathrm{mod}\big( s_n + t\times BW, 2^{SF}\big)\times \frac{BW}{2^{SF}} -\frac{BW}{2},
        \end{equation}
        and $\mathrm{mod}(\xi, 2^{SF})$ is the remainder of the division of $\xi$ by $2^{SF}$. 


    In LoRa $SF \in \{7,8,9,10,11,12\}$. It is customary to represent a chirp in discrete-time using $2^{SF} \times f_s/BW$ samples indexed by $k$, where $f_s$ is the sampling frequency and $f_s/BW$ is the oversampling factor. Letting $t=k/f_s$, we obtain:
    \begin{multline} \label{eqn:lora_mod}
        c_n (k) = \frac{1}{\sqrt{2^{SF}}}\exp\bigg\{j2\pi\times \Big(f_c+ \\ \mod\big(s_n+ k \times \frac{BW}{f_s}, 2^{SF}\big)\times \frac{BW}{2^{SF}}-\frac{BW}{2}\Big)\bigg\} \times k,
    \end{multline}
    where $k = \{0, 1, 2, \ldots, (2^{SF}\times f_s/BW)-1\}.$ 
 \Cref{fig:chirp} shows a chirp in continuous time, in discrete time and in its time-frequency representation.


    \section{System description}
    
            Traditionally, jamming in the physical layer corresponds to adding white Gaussian noise (AWGN) to the transmitted signal. Such na\"ive strategies are ineffective in LoRa communications.
            Due to that resiliency to AWGN, LoRa has also been referred to as a secure communication protocol. However, it has been shown by \cite{hou2023jamming} that LoRa is vulnerable to jamming using a chirp-type waveform. 
            Generating the chirp-type waveform to cause destructive interference requires the knowledge of $BW$ and $SF$. 
            
            The LoRaWAN specification fixes the choice of these parameters to a finite set of $18$ combinations ($BW \in \{125\mathrm{kHz}, \ 250 \mathrm{kHz}, \ 500\mathrm{kHz}\}$  and $SF \in \{7, 8, 9, 10, 11, 12\}$). These parameters are agreed by the legitimate communicating parties, but are not readily available to a jamming adversary. Hence, the jammer needs to estimate this information from an observed signal.
            
            \begin{figure}[!b]
                \centering
                \includegraphics[width=\columnwidth]{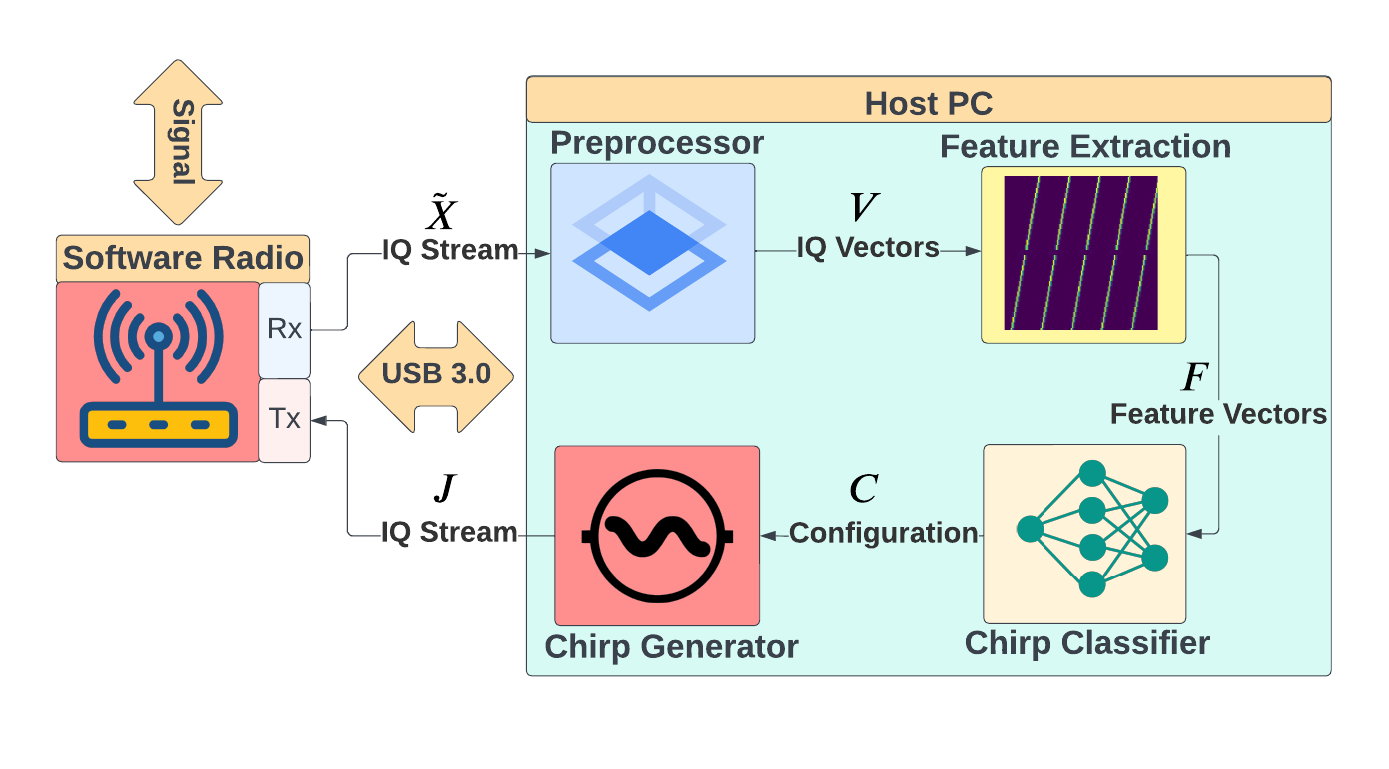}
                \caption{Block diagram for a reactive jammer in a communication system that uses CSS modulation.}
                \label{fig:sys_model}
            \end{figure}

            \Cref{fig:sys_model} shows the block diagram of the data pipeline used by a reactive LoRa jammer. Each component of this system is described in the following subsections.

        \subsection{Data batch preprocessing block}

            The SDR captures signals in real time and outputs a stream of In-phase and Quadrature (IQ) samples of indefinite length. On the other hand, our neural network classifier operates on data batches of finite size. The \textit{preprocessor} block collects data flowing in from the SDR into a matrix of appropriate size for processing in the subsequent blocks. 
            
            The SDR is tuned to the channel of interest and configured to a sampling rate of $1\mathrm{MHz}$. Due to the Shannon-Nyquist Theorem, a minimum sampling rate of $1\mathrm{MHz}$ is required since the maximum $BW$ in LoRa is $500\mathrm{KHz}$. A lower sampling rate might result in distortion from aliasing, and higher rates imply higher demand for computational resources.
            Therefore, the SDR generates a noisy IQ stream $\tilde{X}$ of discrete-time samples to the host PC. The preprocessor block parses this stream of complex values into smaller signal blocks and reshapes them into a matrix of dimensions $B \times M$. Where $B$ represents batch size and $M$ represents length of the signal segment. 
            


            Determining the proper block length $M$ is crucial, as it must contain enough samples to distinguish the LoRa configurations reliably. If the block length is too small, the signal is truncated and information is lost. If the block length is too large, the the neural network processing introduces latency. Hence it must be as small as possible yet carry enough signal information. 

            We have empirically determined that the ideal block length must span two LoRa symbols for the longest configuration. The longest configuration in LoRa is  $BW = 125\mathrm{KHz}$, and $SF = 12$, resulting in a symbol duration of $T_s = 2\times2^{12}/125000$ seconds. For a sampling frequency of $1\mathrm{MHz}$, we obtain an over-sampling factor of $8$, resulting in $2\times2^{12}\times8=65,536$  of samples. Therefore, we fix the block length to $M=65,600$.
            
        \subsection{Feature Extraction}

            The \textit{feature extraction} block employs an algorithm based on the instantaneous frequency (IF), which leads to a compact representation of LoRa signal sequences. Such representation  accentuates features related to the identification of $BW$ and $SF$. The algorithm first transforms the signal vectors from the time domain to the frequency domain and tracks the instantaneous frequency of the signal over time. In the frequency domain, any pair of LoRa signals corresponding to different configurations appear distinctly different. The algorithm takes in a batch of signal blocks from the preprocessor block, $V$,  and applies the algorithm described in \Cref{se:feature_extraction} to produce a matrix $F$ of IF vectors.
            
            Our goal is to infer the parameters $SF$ and $BW$. One influences the duration of the chirp, and the other affects both the duration and frequency sweep range in the chirp. Our approach to feature extraction here is to characterize the instantaneous frequency of the signal, describing the evolution of the frequency in the signal with time. Through this representation, we can observe both the range of the frequencies swept and the time elapsed for each sweep, enabling simultaneous estimation of $SF$ and $BW$.

            
        \subsection{Chirp classifier}

            The \textit{chirp classifier} block uses a neural network (NN) to identify the transmitted chirp signal. Our model is trained using a dataset of IF vectors labeled with their corresponding $BW$ and $SF$ configurations. 
            Once trained, this block receives an IF vector and performs a \textit{soft-decision} classification of $BW$ and $SF$ in a vector $C$ of probabilities for each of the $18$ possible signal configurations. 
            This information passed to the chirp-generator block.



In the context of classifying LoRa signals based on their features, it is important to note that the relationship between these features and their respective classifications is non-linear. NNs can learn complex relationships and patterns in data, making them suitable for tasks like classifying signals with intricate or non-linear relationships between their features and categories.
With proper training and a sufficiently rich architecture, NNs can provide accurate signal classification even at extremely low levels of SNR. We will discuss the NN architecture in more detail on~\Cref{se:neural_network}.
            

        \subsection{Chirp generator}

            
            The \textit{chirp generator} block is responsible for utilizing the inferred $BW$ and $SF$ to generate a stream of discrete-time IQ values for the jamming chirps, denoted by $J$. The IQ stream should be sent to the SDR, which uses a Digital to Analog Converter (DAC) that converts them from discrete-time to a corresponding continuous-time signal. Once converted to analog, the SDR can adjust the signal to the channel's center frequency for transmission. The resulting signal would represent a chirp with the same $BW$ and $SF$ as the target signal, leading to interference at the receiver.
            

            LoRa uses a two-step demodulation procedure: the first is known as \textit{dechirping}, followed by an FFT. The dechirping operation multiplies the sampled signal with a base down chirp of the same $BW$ and $SF$. The resulting signal has a constant frequency, which matches the chirp's initial frequency. Then, from its FFT, we identify the bin index of this frequency, determining the encoded symbol's value. Under this demodulation scheme, when two signals of the same $BW$ and $SF$ configuration interfere at the receiver, they result in multiple indiscernible peaks in the FFT step. Such interference deceives the receiver into misidentifying the original symbol. This misidentification leads to symbol demodulation errors, resulting in packet drops, effectively jamming the signal.
            

            With the knowledge of $BW$ and $SF$ we can generate chirp signals using \cref{eqn:lora_mod}. However, the chirp's polarity (\textit{upchirp} or \textit{downchirp}), the symbol value, and the arrival time influence the effectiveness of interference with the target signal. Considering these factors, the authors of~\cite{hou2023jamming} introduced three effective methods to jam LoRa signals when $BW$ and $SF$ are known, which can be implemented in the chirp generator block, summarized as follows:
            
            \begin{itemize}
                \item \textbf{Identical chirps:} A simple approach is to continuously repeat the same symbol in sequence. By transmitting continuously, we avoid sudden shifts across demodulation windows. Any delays and time offsets only affect the initial frequency of the chirp and still result in demodulation errors. This method is lightweight because it does not require strict time synchronization.

                \item \textbf{Consecutive downchirps:} This method targets the Start Frame Delimiter (SFD) symbol of LoRa packets, which  is a base downchirp that marks the beginning of the packet header. From transmitting base downchirps consecutively, the receiver is tricked into making errors in identifying the legitimate SFD, resulting in incorrect packet parsing and leading to packet drops.
                
                \item \textbf{Synchronized chirps:} This method is considered to be the most effective jamming strategy in LoRa \cite{hou2021jamming,hou2023jamming}. It involves transmitting random symbols that perfectly align with the demodulation window at a receiver. This is made possible by estimating and compensating the Carrier Frequency Offset (CFO) and the Sampling Time Offset (STO),  as in a legitmate LoRa demodulator. The 
                \textit{synchronized chirps} method requires strict synchronization and additional computing, however, it is the most effective and difficult to detect method known to date.
            \end{itemize}

            In conjunction with the inferred parameters, the chosen method defines the sequence of jamming chirps to be transmitted. The IQ values corresponding to this sequence is streamed from the chirp generator block to the SDR at a fixed rate. Consequently, the SDR transmits this waveform over the air to jam the target signal at the receiver. This strategy shows that it is possible to jam LoRa signals of  unknown $BW$ and $SF$ configurations by an unauthenticated agent.
                    
            
    \begin{figure*}[!htb]
        \centering
        \includegraphics[width=0.9\textwidth]{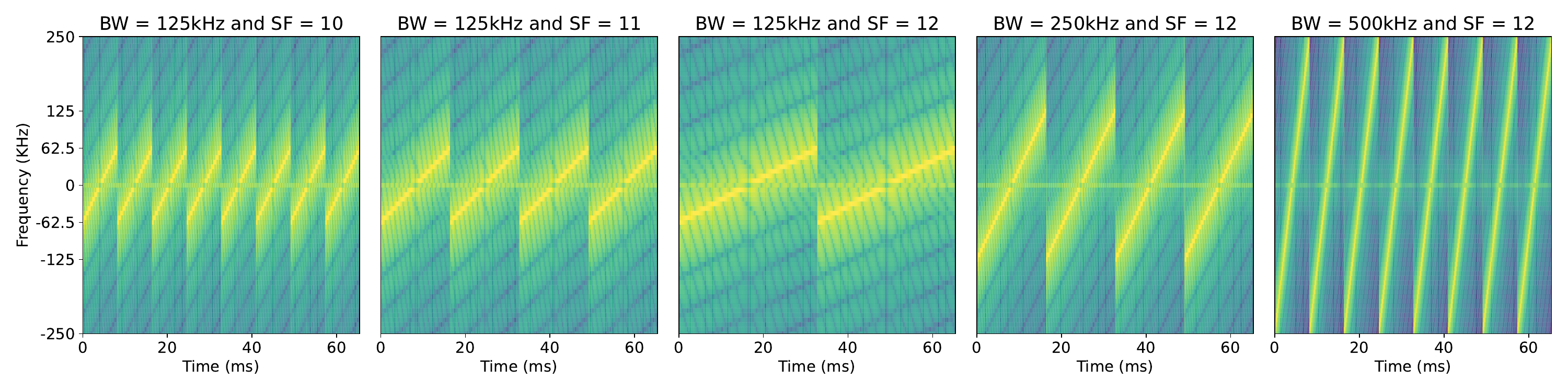}
        \caption{Feature representations for various LoRa Configurations}
        \label{fig:feature_plots}
    \end{figure*}

    
    \section{Feature extraction} \label{se:feature_extraction}

        
        In this section, we identify a pattern in the data, also known as feature, that aids in distinguishing one category from another. To that end, we compute the \textit{instantaneous frequency} of the signal. Considering this feature, we can retain information about the range of frequencies swept, and their sweep rate simultaneously, which are directly related to our two parameters of interest, $BW$ and $SF$.
        
        Here, we follow a two-step procedure to computing the instantaneous frequency: a Short Term Fourier Transform (STFT) followed by Instantaneous Frequency (IF) estimation.    
        
        \subsection{Short Term Fourier Transform (STFT)}

            Given the inherent time-varying nature of frequency in a chirp signal, we employ the STFT on each input signal segment \cite{allen1977short}. A given signal segment is further subdivided into overlapping windows, each consisting of $W=128$ samples, with an overlap of $L=64$ samples. 
            Subsequently, an FFT is executed on these windows. This operation obtains the power distribution across all the frequencies in the channel bandwidth, $BW$ as the signal evolves in time, as follows:
            \begin{equation} \label{eq:stft}
                Q[k,m] = \sum_{n=0}^{W-1} x[n+mL]w[n]e^{-j2\pi nk/W},
            \end{equation}
            where $Q[k,m]$ is the STFT coefficient at frequency bin $k$ and time index $m$, $x$ is the input signal segment, and $w[n]$ is the Hann window function \cite{testa2004processing}. 
            
        \subsection{Instantaneous Frequency Estimation}
            
            Unlike stationary signals where the spectral properties are constant, the frequency of a chirp signal varies linearly with time \cite{boashash1992estimating}. For such signals, we must compute the \textit{instantaneous frequency} instead of frequency. The instantaneous frequency is a time-varying parameter related to the average of the frequencies present in the signal as it evolves in time~\cite{boashash1992estimating2}. 
            
            From the STFT operation in \cref{eq:stft}, we obtain the energy distribution over all frequency bins for every time-step. We use this energy distribution to compute a weighted average of the frequencies at each time-step, obtaining the instantaneous frequency of the signal, as follows:
            \begin{equation} \label{eq:inst_freq}
                f_{inst}(m) = \frac{\sum_{k=1}^K P(k, m)f(k, m)}{\sum_{k=1}^K P(k, m)},
            \end{equation}
            where $f_{inst}(m)$ is the instantaneous frequency at the time index $m$, $f(k, m)$ is the peak frequency at frequency index $k$ and time index $m$, and $P(k, m)$ is the power spectral density, computed as $P(k, m) = |Q[k,m]|^2$.
            

            
        
    \section{Neural network architecture} \label{se:neural_network}

        LoRa nodes operate under power constraints (typically from $10\mathrm{dBm}$ to $20\mathrm{dBm}$) and often transmit over long communication distances (typically from $10^3$m to $10^4$m). As a result, LoRa signals are often received at low SNR, sometimes even below the noise floor. Identifying and distinguishing such signals reliably demand a classifier model with high noise tolerance and discriminative power.
            
        Neural networks have been extensively used for signal classification in wireless communications, spanning applications such as channel sensing, interference detection and spectrum management \cite{fehske2005new, oveis2023convolutional, si2023efficient}. Central to their efficacy in these applications is their inherent ability to model non-linear relationships between parameters and noisy data \cite{sharma2017activation}. 
        
        
        \Cref{fig:feature_plots} illustrates the feature representations corresponding to different $BW$ and $SF$ configurations. The first three sub-figures show the case of fixed $BW$, and the last three figures illustrate the case of fixed $SF$. These graphs indicate that changes in $BW$ and $SF$ result in clearly distinct waveforms. Additionally, the characterization based on the IF of these waveforms makes the task of distinguishing signals of different configurations much simpler by converting the the problem of estimating $BW$ and $SF$ into a signal classification problem. 
        
\begin{figure}[b]
            \centering
            \includegraphics[width=0.85\columnwidth]{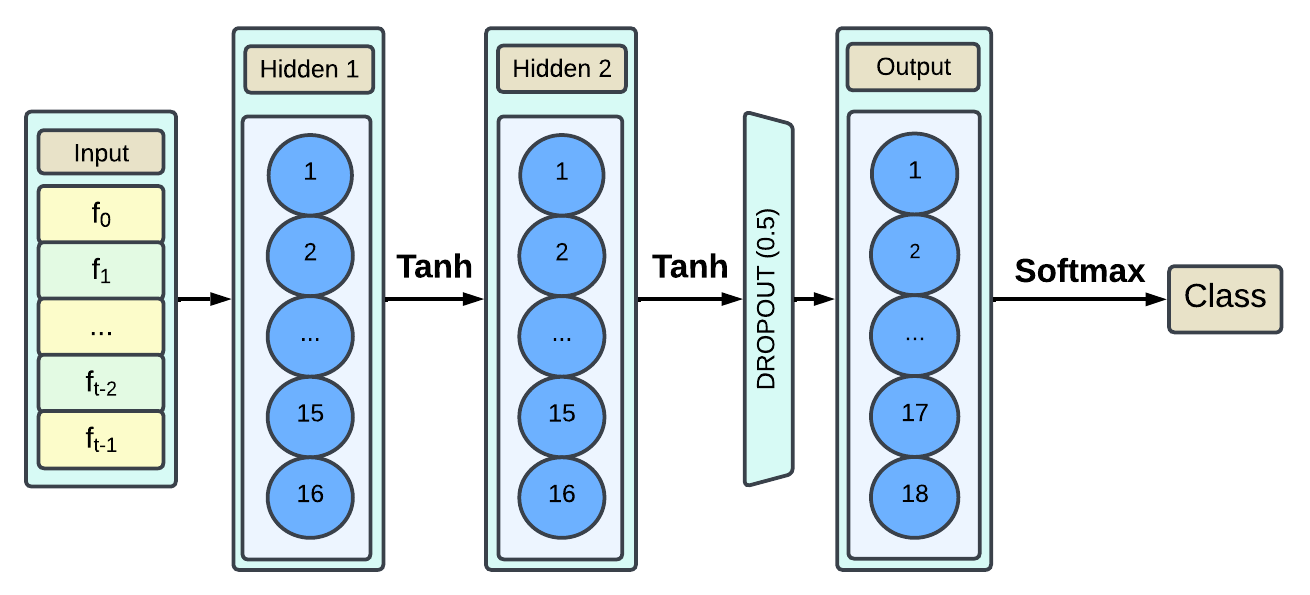}
            \caption{Neural network architecture used in our system.}
            \label{fig:neural_net}
        \end{figure}

        For this classification task, we use a feed-forward neural network as illustrated in \Cref{fig:neural_net}. The model features two hidden layers with 16 neurons each, and an output layer of 18 neurons, as specified in \Cref{table:nn_attributes}. The input is the IF vector, where $t$ is the time index. To classify an IF vector into one of the $18$ categories, we use a softmax function in the output layer to obtain a probability distribution on the likelihood of each class given the observed data. Consider an output of the final layer, $Z = [z_1, z_2, \dots, z_{18}] $ of $18$ real numbers, the softmax function, $S(\cdot)$, is defined as:
        \begin{equation} \label{eq:softmax}
            S(z_i) = \frac{e^{z_i}}{\sum_{j=1}^{18} e^{z_j}}, \ \ i = 1, \dots, 18.
        \end{equation}

        \begin{table}[tb]
        
            \centering
            \caption{Key Attributes of the Neural Network Architecture}
            \begin{tabular}{|c|c|}
            \hline
            
            \textbf{Attribute} & \textbf{Description} \\
            \hline
            \hline
            Input & Flatten Layer \\
            \hline
            Hidden Layer 1 & Dense (16 units, tanh activation) \\
            \hline
            Hidden Layer 2 & Dense (16 units, tanh activation) \\
            \hline
            Regularization & Dropout (0.5 rate) \\
            \hline
            Output Layer & Dense (18 units, softmax activation) \\
            \hline
            Loss Function & Categorical Cross-Entropy \\
            \hline
            Optimizer & Adam \\
            \hline
            Evaluation Metric & Classification Accuracy \\
            \hline
            \end{tabular}
            \vspace{3pt}
            
            \label{table:nn_attributes}
        \end{table}

    \section{Simulation Results} \label{se:evaluation}
    
        We use synthetic datasets of LoRa signals, creating separate datasets for training and validation.\footnote{The data and code for all the simulations and numerical experiments in this paper are available at \url{https://github.com/MINDS-code/jammingSDR.git}}. Here, the noisy signals are generated according to an Additive White Gaussian Noise (AWGN) model producing signal data at diverse SNR levels.  Our training dataset has $10$ SNR levels, ranging from $0$ to $20 \mathrm{dB}$. For each of the $18$ configurations, we have generated $50$ signal files. Thus, the training dataset contains a total of $9000$ entries. Our validation dataset has a broader SNR range, from $-15$ to $20 \mathrm{dB}$, leading to $18$ SNR levels in total. Here, we have generated $20$ signal files for each case, leading to a total of $6480$ entries. We found out that if the training dataset included signals with SNR below zero, the classification performance of the of the NN is severely degraded. 
        
        \begin{figure*}[!ht]
            \centering
            \includegraphics[width=0.6\columnwidth]{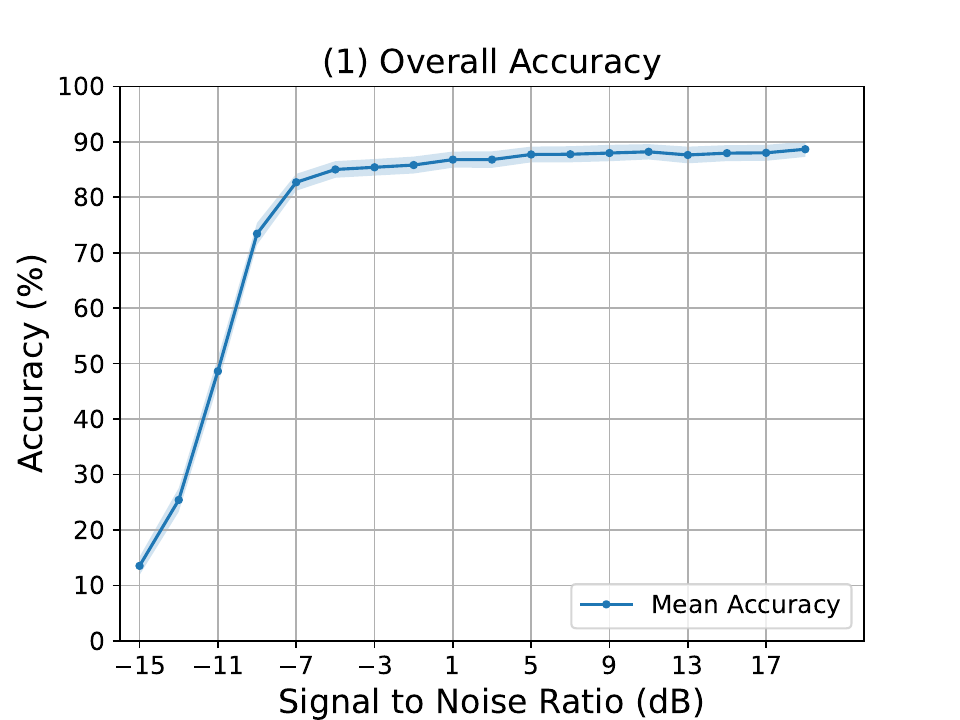}
            \includegraphics[width=0.6\columnwidth]{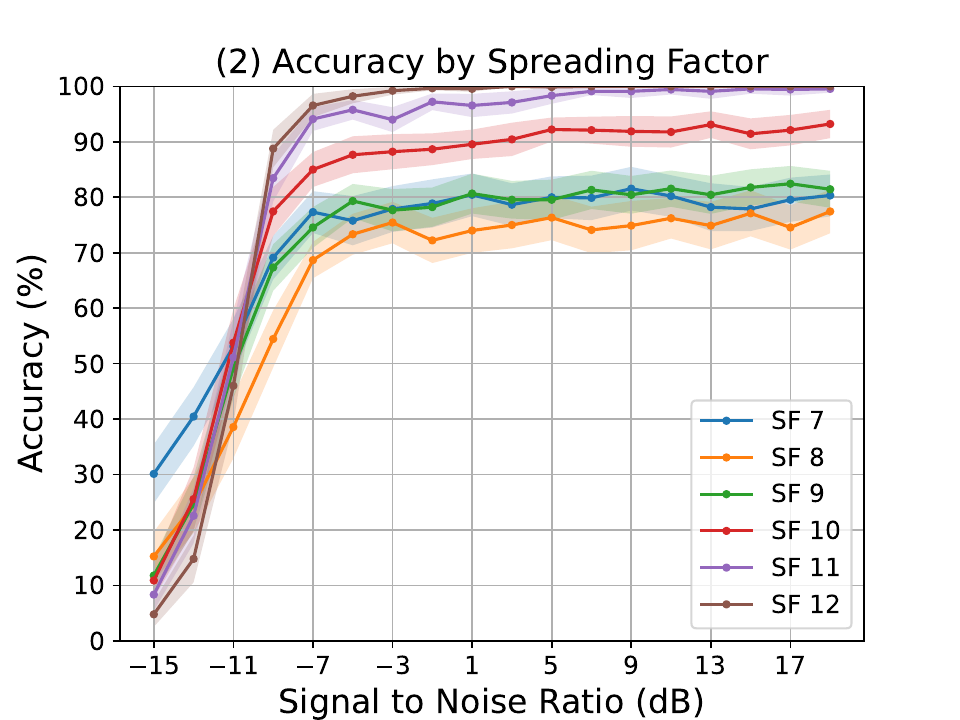}
            \includegraphics[width=0.6\columnwidth]{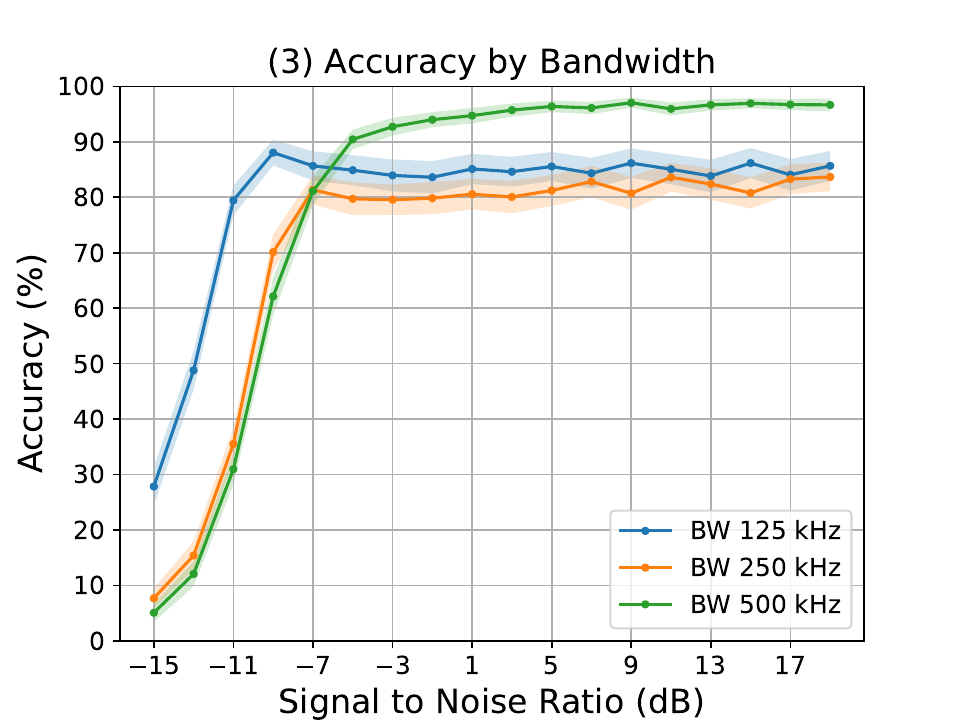}
            \caption{Classification accuracies against SNR with a 95\% confidence interval: (1) overall, (2) by spreading factor, and (3) by bandwidth.}
            \label{fig:classification_accuracies}
        \end{figure*}        
        Consider a clean signal, denoted by $X$, subjected to AWGN denoted by $Z$ as follows:
        \begin{equation}
            \label{eq:additive_noise}
            \tilde{X} = X + Z,
        \end{equation}
        where $\tilde{X}$ is the resulting noisy signal. The power level of $Z$ is determined by the desired SNR level.        
        We obtain confidence intervals on results by repeating the experiment $30$ times. Addionally, we experiment with fixed $BW$ and $SF$ choices, observing their influence on classification performance. 
        \Cref{fig:classification_accuracies} (left) illustrates the classifier's overall accuracy in relation with SNR. Classification accuracy starts at around 12\% for $-15 \mathrm{dB}$ SNR. The accuracy increases sharply and saturates at $-5 \mathrm{dB}$ SNR.
\Cref{fig:classification_accuracies} (middle) illustrates the classifier's accuracy as a function of SNR for the different possible $SF$ configurations. Each curve differs from the others by their saturation points and the accuracy levels they can reach. The curve for $SF$ 12 reaches saturation the earliest and at the highest accuracy level, succeeded by $SF$ 11, with subsequent configurations following in descending order. 
        We observe that for a fixed SNR level, higher SF choices yield consistently higher accuracy scores. The mean classification accuracy improved with an increase in $SF$ from 7 to 12. This trend results from LoRa's spreading waveform, where $SF$ determines the sweep rate of the chirp. A higher $SF$ leads to a longer chirp duration resulting in a more elongated and discernible frequency trajectory over time. With more samples constituting the waveform, identification becomes more precise, improving classification accuracy. 
        
        \Cref{fig:classification_accuracies} (right) illustrates the classifier's accuracy as a function of SNR for three $BW$ configurations: $125 \mathrm{KHz}$, $250\mathrm{KHz}$, and $500\mathrm{KHz}$. Before reaching saturation, the $125 \mathrm{KHz}$ curve shows a higher accuracy compared to the other two. Meanwhile, the classifier's accuracy for the $250\mathrm{KHz}$ curve is consistenlty higher than for $500\mathrm{KHz}$.
        
        After reaching saturation, the $500\mathrm{KHz}$ curve exhibits higher accuracy over the other two. However, beyond this point, all three configurations deliver high classification accuracy. Thus, despite the high accuracy of the $500\mathrm{KHz}$ curve post saturation, the real differentiator lies in their points of saturation. The earlier the saturation, the lower the minimum SNR needed to classify the signal reliably. Thus, the order in which the curves saturate imply that lower $BW$ configurations yield better detection. 
        
        The mean classification accuracy saturates later for higher $BW$ choices from $125\mathrm{KHz}$ to $500\mathrm{KHz}$. With a wider $BW$, the signal's frequency changes on a broader range in a reduced period. This rapid shifting causes the instantaneous frequency vectors to become too closely spaced, making it more challenging for the classifier to distinguish them.

        The choice of $BW$ and $SF$ in LoRa is motivated by the application's quality of service requirements. However, in practice, nodes switch between several parameter choices to save power and optimize throughput. Therefore, when jamming or extensive interference is a concern, legitimate nodes must consider switching to faster $BW$ and $SF$ choices. Our results conclude that, to avoid detection by unauthorized agents, legitmate LoRa nodes must opt for lower $SF$ choices and higher $BW$ choices whenever possible.

        
    \section{Conclusions and future work}

        Many large-scale multi-agent systems rely on LPWAN protocols. Amongst these, LoRaWAN has found widespread adoption, due to its energy efficiency, long range, and use of unlicensed spectrum. However, it is succeptible to cyber-attacks, including eavesdropping and jamming. In this paper, we explored the vulnerability of LoRa to signal jamming.

        A survey of related literature revealed that LoRa is vulnerable to jamming with a particular chirp type signal. However, generating such signals require the knowledge of the bandwidth, and spreading factor of the target LoRa signal. We argue that this information is shared amongst legitimate parties but unavailable to an unautheticated adversarial agent. In this work, we presented the high-level design of a practical jammer, that makes use of a neural network classifier for estimating these parameters by eavesdropping and reactively emits jamming chirps.

        Leveraging a structural pattern in LoRa’s signal waveform, we relate the problem of estimating these parameters to a signal classification task. To that end, we proposed a feature extraction method that computes the instantaneous frequency of signals, enhancing features pertinent to identifying $BW$ and $SF$ configurations. Then we trained a feedforward neural network classifier on a dataset LoRa signals to learn these characteristics for predictive analysis. Our results indicate that the classifier begins to reliably estimate these parameters for signals stronger than $-5 \mathrm{dB}$ SNR. Additionally, we analyzed detection performance at various configurations of $BW$ and $SF$. Ultimately revealing that, to hinder such detection, legitimate users of LoRa must use lower $SF$ and higher $BW$. 
        
        Directions for future work include experimenting this classifier on a dataset of real signals captured using a software radio to provide a real-world validation of this analysis, and an end-to-end implementation of the proposed jammer to explore real-time performance of the design.
        
        

\bibliographystyle{IEEEtran}
\bibliography{IEEEabrv,References}
    
\end{document}